\documentclass[11pt,a4paper,twocolumn,USenglish]{article}

\usepackage[top=.7in,left=.7in,right=.7in,bottom=1.1in]{geometry}
\usepackage{setspace}

\usepackage{fourier}
\usepackage[USenglish]{babel}
\usepackage[utf8]{inputenc}
\usepackage{authblk}

\usepackage{multirow}
\usepackage{natbib}
\usepackage{tabularx}
\usepackage{booktabs}
\usepackage[export]{adjustbox}
\usepackage{subfig}
\usepackage{caption}
\usepackage{url}
\usepackage[bookmarks, colorlinks, breaklinks]{hyperref}
\usepackage[usenames,dvipsnames,table]{xcolor}
\usepackage{colortbl}

\usepackage{outlines}
\usepackage[linewidth=0.5pt]{mdframed}
\definecolor{LightCyan}{rgb}{0.88,1,1}
\definecolor{MyGray}{gray}{0.9}
\usepackage{subfig}

\usepackage{booktabs}

\usepackage{rotating}

\title{Computational appraisal of gender representativeness\\ in popular movies}

\author[1]{Antoine Mazi\`eres\thanks{Corresponding author: antoine.mazieres@gmail.com}}
\author[1]{Telmo Menezes}
\author[1,2]{Camille Roth}
\affil[1]{CNRS, Centre Marc Bloch, \href{https://cmb.huma-num.fr}{Computational Social Science team}, Berlin, Germany}
\affil[2]{CAMS, Centre d'Analyse et de Math\'ematique Sociales, CNRS/EHESS, Paris, France}

\date{}

\begin{document}

\newcommand{\ffr}{\textsc{ffr}}

\newcommand{\tb}[1]{\textcolor{blue}{#1}}
\newcommand{\m}[1]{\tb{#1}}
\newcommand{\apr}[1]{\textcolor{Apricot}{#1}}
\newcommand{\todo}[1]{{\sc\apr{[#1]}}}
\hypersetup{linkcolor= MidnightBlue,citecolor= MidnightBlue,filecolor=black,urlcolor= MidnightBlue}
\definecolor{gray}{rgb}{0.98,.98,.98}
\definecolor{darkgray}{rgb}{0.8,.8,.8}
\definecolor{darkdarkgray}{rgb}{0.66,.66,.66}

\newcommand{\tg}[1]{\textcolor{green}{#1}}
\newcommand{\g}[1]{\tg{#1}}
\newcommand{\tr}[1]{\textcolor{red}{#1}}
\renewcommand{\r}[1]{\tr{#1}}

\maketitle

\begin{abstract}

Gender representation in mass media has long been {mainly} studied by qualitatively analyzing content. This article illustrates how automated computational methods may be used in this context to scale up such empirical observations and increase their resolution and significance. We specifically apply a face and gender detection algorithm on a broad set of popular movies spanning more than three decades to carry out a large-scale appraisal of the on-screen presence of women and men. Beyond the confirmation of a strong under-representation of women, we exhibit a clear temporal trend towards a fairer representativeness. We further contrast our findings with respect to movie genre, budget, and various audience-related features such as movie gross and user ratings. We lastly propose a fine description of significant asymmetries in the \emph{mise-en-scène} and \emph{mise-en-cadre} of characters in relation to their gender and the spatial composition of a given frame.

\vspace{1em}
\vspace{1em}

\end{abstract}

\section*{Introduction}

There is assuredly a long tradition of scholarship in the description of sex roles on mass media of various types: already in her seminal review, Linda \cite{busby1975sex} described how instructional material, TV, films, advertising, newspaper, cartoons and literature have been used since the late 1950s to study gender-related representations such as sexual stereotypes, biases in occupational roles, body staging, marriage and rape. Back then, she further concluded that ``media sex-role studies that have been completed in the 1960s and early 1970s can be used as historical documents to measure future social changes'', emphasizing the need of replicating these analyses at several points in time to capture underlying mutations and trends. As empirical material, such sources provide the opportunity to grasp a certain state of affairs regarding gender representations, together with the intents and conflicts of interest at play in shaping them. Recent reviews of this research \citep{rudy2010context,collins2011content} highlight the ubiquity of gender patterns, most notably the under-representation and sexualization of women, across multiple media and content types, even though some negative results may occasionally be found as well \citep{kian-espn-2009}. Almost a half century after Busby's review, the roles of females and males in media and fiction have been a prominent domain of inquiry in content analysis and have been subjected to many analyses {based on a sometimes substantial quantity of cultural artifacts} \citep{neuendorf2016content}, including for instance broadcast network programs \citep{lauzen2018boxed}, popular movies \citep{lauzen2019sa,smith2019inequality} and recurring TV show characters \citep{glaad2019tv}.

Methodologically, this {strand of media gender} research principally relies upon {manual} assessments of text, images and scripts, which {occasionally} feature complex semantic concepts and possibly subjective interpretations. As a result, these approaches are difficult to scale to a large number of observations{: a lot of human coders are required to perform statistical and especially temporal analyses}.
Some studies do rely on large-scale and automatically collated datasets, for instance through collaborative platforms  such as IMDb, the Internet Movie Database, but they are by definition limited to {already-available} metadata,
such as film cast, crew, or budget \citep{lindner2015million,yang2020measuring}.
{The systematic} construction and extraction of
variables adequate for a given study and a given research question remains a challenge. 

Recent advances in artificial intelligence and data science may significantly help in this regard, especially in terms of automated processing of text, image and video, where current technologies are sometimes capable of competing with humans in a wide array of specialized tasks, including automatic text summarization \citep{mani-auto}, topic detection \citep{chaney2012visualizing}, or translation \citep{hassan2018achieving}; face recognition \citep{dhomne2018gender,guo2019survey}, scene intensity estimation \citep{Kataria2016SceneIE}, narrative element extraction \citep{guha2015computationally,bost2016narrative}; or even at the interface of both, text description generation from images \citep{xu2015show}. At the moment, however, these methods have generally been applied on issues that remain quite close to the scientific fields which they originate {from: they focus rather on technological than social science applications.}

Our contribution explores the possibility of using such advances to the construction of datasets relevant to sex role research. Firstly, we outline a field of inquiry by focusing on cinema, for which we identify a relevant subset of {more that $3\,500$ popular movies spanning over 3 decades}. We extract a representative set of frames from this dataset and applied machine learning models to detect human faces and infer their gender. We take the extra precaution of evaluating the performance and fairness of these inferences regarding the target categories (\emph{female} and \emph{male}), for these models are typically evaluated in all generality and their potential biases may vary with respect to data corpora. Secondly, we devise a metric to appraise women's presence in movies, the \emph{female face ratio} (\textsc{ffr}). We compare it with another well-established measure, the Bechdel test. In aggregate, \ffr{} markedly increases over time, to the point of approaching female-male parity. Also, there are significant differences in how its values are distributed for successive temporal periods. This indicates a noticeable mutation in the popular movie-making culture regarding women's representation. Thirdly, we explore several more sophisticated and experimental capabilities of automatic face detection to analyze how characters of distinct genders are framed on-screen. Interestingly, this yields mostly negative results in the sense that we observe very little variations. We nevertheless exhibit a few significant patterns related to gender-mixed environments.

{A few recent academic endeavors have started exploring methodologies of automated visual content analysis in a social science framework.
These works have been denoted with a variety of labels.
In the context of digital humanities, for instance, the notion of ``distant viewing''~\citep{arnold2019distant} has been coined} by analogy with the famous concept of ``distant reading''~\citep{moretti-2000-conjectures}. {The emerging field of so-called ``computational media intelligence''~\citep{somandepalli2021computational} covers a variety of initiatives with a more technical focus~\citep{guha2015computationally,Kataria2016SceneIE}. In this area, a case study aimed at tracking female participation in the 100 top-grossing Hollywood films over 6 years is notably relevant here \citep{guha2015gender,somandepalli2021computational}, as it introduced algorithms specifically designed to measure on-screen presence and gender-specific speaking time.
In a similar vein, \cite{jang2019quantification} applied an object detection system on 900 movies to characterize which items were present in association with a face of a given gender, and how often.}

{Our research belongs to this strand. On the one hand, we rely on a relatively simple and mainstream algorithmic apparatus enabling face detection and gender inference from still frames. In this regard, our contribution is more methodological than technical: we focus particularly on the construction of a sound protocol that pays special attention to a form of criticism prevalent in social sciences regarding the potential biases induced by the use of automated labeling methods, especially when stemming from machine learning approaches~\citep{buolamwini2018gender,crawford2019excavating}.
On the other hand, we apply our method on a much larger dataset than has been done so far, and on a much wider period of time. This enables us to originally analyze the temporal evolution of gender representativeness in films over decades.} 

More broadly, we contend that the systematic application of such techniques could {contribute} to the formulation of ambitious research questions that would be hardly tractable with {only a} human workforce. This could furthermore enable the creation of well-documented datasets featuring metadata adapted to sex role research for the community{, in order} to thoroughly and conveniently reproduce and improve experiments. Tackling this challenge could indeed trigger new fields of interest {for both qualitative and quantitative approaches. For} instance, {this could help} formulating a theoretical understanding of the distribution of representations over the whole spectrum of a specific medium, or focusing on potential outliers
in order to unveil their possible contribution to {future evolutions}.

\section*{Dataset and data processing}

\subsection*{Corpus scope}

Movie studies typically define the corpus scope by relying on box office data as a proxy for movie popularity~\citep[e.g.,][]{follows2014gender,lauzen2019sa,smith2019inequality}.  They essentially outline a selection based on the yearly top grossing movies over a period of time \hbox{i.e.,} short-term commercial success in movie theaters, which is admittedly related to popularity. {Yet} popularity relies on complex behaviors: it relates as much to the value given by an individual to the content, as to the value an individual perceives, or anticipates, others will give. Intricate interactions of support, rejection, controversy, advocacy and imitation come into play to establish a cultural object's influence~\citep{cillessen2011conceptualizing}. Put shortly, attendance alone may not help fully capture movies that are both characteristic of cultural representations and influential in shaping them.
{In particular, it may discard some content that may qualify as ``mainstream'' yet did not attain significant box office success.}

We thus devised a different approach based on open collaborative platforms such as peer-to-peer file sharing networks~\citep{vassileva2002motivating,cohen2003incentives} or wiki-based knowledge sharing systems~\citep{rafaeli2008online,yang2010motivations}. {These online environments} are fueled by interactions between a diverse and critical mass of users. Contributors are incentivized by the effort of others to increase the system usefulness by creating and maintaining fashionable resources: they act from a variety of motives, including both the perceived value of the content they provide and the peer recognition that it entails. We argue that the intensity of such collaborative activity {defines a broader proxy of content mainstreamness than attendance. However, we also acknowledge that it may be biased toward the notoriously younger population of such online communities and their tastes.}

Based on this, we focus on films for which data is available on two significantly distinct types of online platforms: (1) a peer-to-peer file sharing network, which is one of the major Torrent communities, YIFY (yts.mx); and (2) a movie-related knowledge-sharing platform, the above-mentioned Internet Movie Database (IMDb, imdb.com), which comprises records on about 500k movies, mostly stemming from user contributions. We first listed all 13,662 movies made available on YIFY, requiring that at least 3 people share them (seeders) as of December 2019. We then linked them to their respective record on IMDb, excluding documentaries and animation movies while requiring that key metadata be available: year of release, genres, users rating, parental rating, runtime, budget and world wide gross.
We find that there are very few movies per year before 1985 (10 on average, no more than 48 for a given year): for the purpose of the {temporal} analysis, we decide to further focus on the period 1985-2019, wherefrom the yearly number of movies per year is always above 100. This yields a dataset of $3,776$ movies. The average runtime is 109 minutes with a standard deviation of 18 minutes, indicating that we essentially gathered feature films. The budget distribution is broad, with a median of \$23m while the first and third quartiles are at \$10m and \$45m, indicating that we focus on a quite diverse array of movie budgets. The same applies to world wide gross figures: median \$43m, first quartile \$11m, third quartile \$122m. This further substantiates our approach for constructing a filter that is broader than when focusing on top audience figures only.

\subsection*{Face recognition and gender estimation}

The computational extraction of artistic or semantic characteristics of a movie traditionally relies on the extraction of a number of significant images~\citep{guha2015computationally,ko2019learning}. This is commonly based on keyframes i.e., frames of a movie's timeline where new shots commence. This method results in better quality images, since keyframes are used as markers for video compression. Also, it likely captures narrative highlights, since a keyframe captures the first state of scenery ---arguably an important one--- from which the shot unfolds. {For one, the previously cited work of \cite{guha2015gender} relied on this approach to downsample movies frames.} However, the duration and pace vary very significantly from a shot to the other, and are also strongly influenced by shot type, movie genre and year of production~\citep{ShotDurations}. Therefore to ensure the representativeness of our sample with respect to what spectators are shown --- even more so for the {temporal} analysis we aim at --- we simply extracted frames on a time frequency basis,  {similarly to what has been done in \cite{jang2019quantification}. Selecting} one image every 2 seconds yielded a collection of more than $12.4$ million images.

We processed each of these images with the help of face detection and gender estimation algorithms provided by a common scientific computing software, \emph{\cite{referencewolfram2020facialfeatures} Mathematica Engine 12}.

We eventually detect close to {$10$} millions faces over more than {$6.6$} million images, with an average of {$2596$} ({$\sigma=1090$}) faces per movie. For every face, the algorithm provides the coordinates of a bounding box, enabling us to take into account both the position and the size of the surface occupied by the face with respect to the frame dimensions. {It also provides an estimation of the likely binary gender of each face (male or female).}

Both algorithms are built using conventional machine learning methods. Many questions have been raised over the recent years regarding the accuracy and potential bias of predictions based on these techniques, and our approach is no exception.
Previous social scientific-oriented research specifically highlighted the issues associated with the construction of the datasets that are used to train machine learning algorithms~\citep{crawford2019excavating}. Put shortly, a dataset of human-labeled pictures is first gathered, such as ImageNet~\citep{deng2009imagenet}. {Labels correspond to categories of interest that should be learned from this dataset, in order to predict them on any unknown dataset. In our case, these labels include} the visible faces (presence and position) and their gender (male or female). Part of {this human-labeled} dataset is {fed to a learning algorithm} ---such as a neural network--- that will initially improvise {predictions} and then, iteratively, learn from its mistakes, readjusting and ultimately converging towards better guesses. The learned model is then tested on another part of the dataset to assess if the algorithm managed to \emph{generalize} well {--- thereby measuring its \emph{performance}}.

Across the state of the art, {both types} of algorithms generally reach accuracies well above 90\%~\citep{guo2019survey,dhomne2018gender}. {Yet, they also display a strong degree of performance variation depending on the type of dataset at hand and, plausibly, the context and type of images, for instance in medical imagery~\citep{zech2018confounding,mcbee2018deep}. Movie frames are likely a specific type of data.} The work of \cite{buolamwini2018gender} on designing \emph{intersectional benchmarks} {is also particularly relevant here, in that it highlights how face detection algorithms perform unevenly} when tested on faces of specific genders or skin tones. {In any event, we thus need to make sure that the algorithms perform sufficiently well with our dataset for our purposes.}

\begin{figure}
	\centering
	\includegraphics[width=\linewidth]{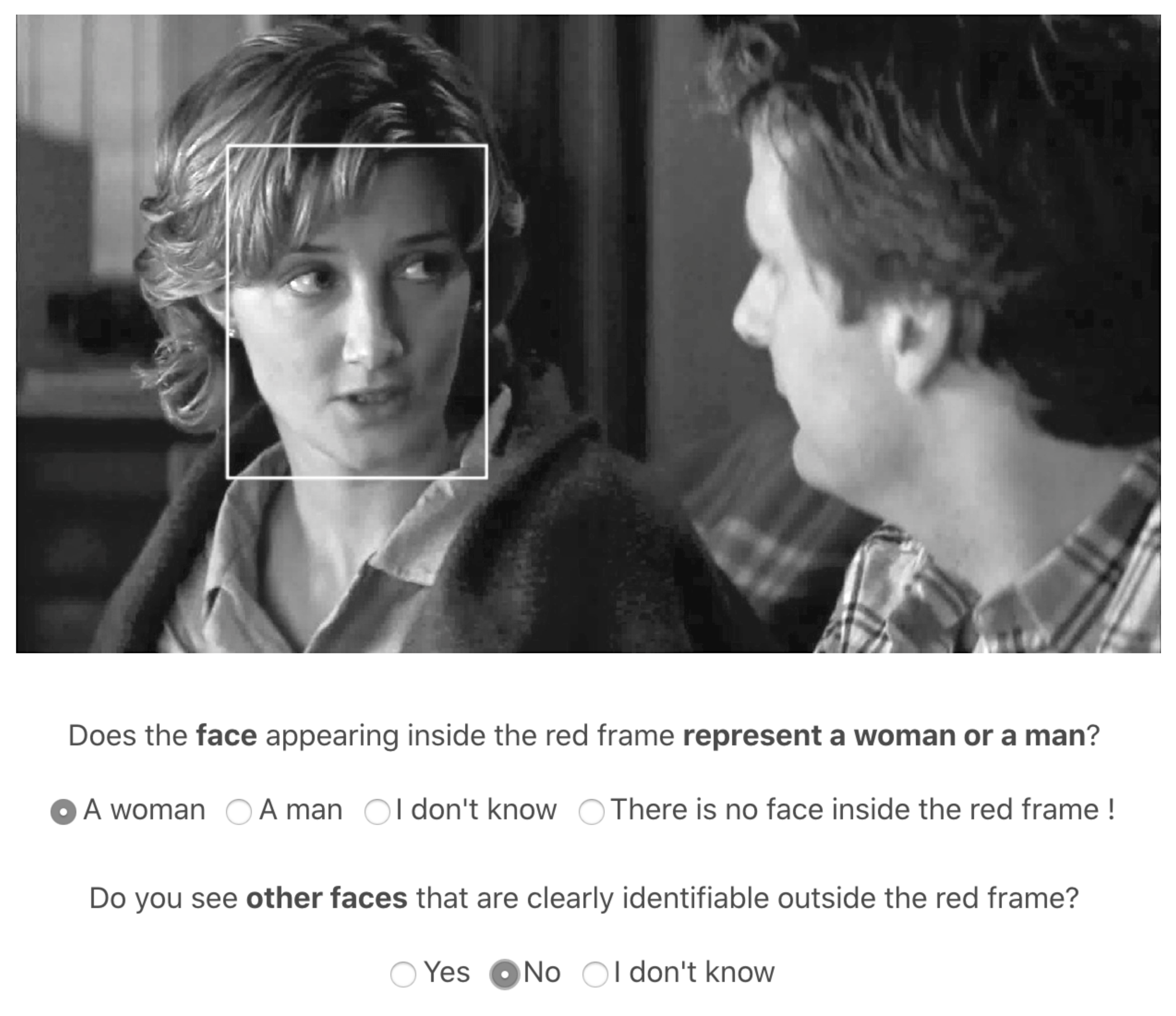}
	\caption{{\bf Interface of the human evaluation experiment}
	}
	\label{fig:screenshot_webapp}
\end{figure}

To this end, we set up a simple experimental protocol:
we randomly select 1000 frames each extracted  from a distinct movie and on which the algorithm detected only one face, half of which female, the other male (so, 500 frames for each gender). We built the web interface shown in Fig.~\ref{fig:screenshot_webapp} displaying one random frame at a time with a bounding box around the detected face, followed by two questions. {The first question} aimed at checking whether the face detected in the bounding box and its gender are correct. {The second question aimed to check whether the frame contains faces outside the bounding box which would therefore be undetected, since only one face was detected on each image.} We sent the link to this website on our research center's internal mailing-list. Participants were invited to review as many frames as they could. Overall, 4,938 reviews were submitted with an average of $4.94$ ($\sigma=2.29$) reviews per frame. For every frame, we considered the most frequent answer. (Narrowing the evaluation only to pictures with identical answers over all reviews actually yielded very similar results). Raw results are gathered on Table~\ref{tab:eval_models}. {For each image, Table~\ref{tab:eval_face_detect} gathers two observations, one for inside the bounding box (true and false positives) and one for the rest of the frame (true and false negatives), thus totaling 2000 observations from 1000 images.}

For face detection, there are 977+863=1840 correct inferences (true positives and true negatives) and 23+137=160 incorrect inferences (false positives and false negatives), thus a high accuracy of 92\%, consistent with the literature. Note that there are much more false negatives than false positives \hbox{i.e.,} the algorithm, when wrong, tends to rather fail to identify a face than erroneously detect one.  Accuracy for gender inference is weaker, with {304+410=714} correct inferences and 162+75+7+8=252 incorrect ones (discarding the negligible ``doubt'' category which indicates that human participants were unable to be conclusive) i.e., a lower yet pretty high 73.9\% accuracy. However, we also notice that gender inference performs quite differently between males and females. When it infers a female face, the face is actually of a women only 65\% of the time, while of a man 35\% of the time. Male faces are accurately identified $84.5\%$ of the time, and are actually of a female for only $15.5\%$ of the cases. 

Therefore, the model shows in aggregate a tendency to wrongly categorize faces as female more often than for male faces. 
It generally informs us that the \emph{raw} inferences of woman faces and thus woman presence are overestimated by the machine learning algorithm that we used.  {While it is clear that a 65\% accuracy in general would be problematic, we luckily deal here with a dichotomized variable: either female or male. Since the accuracy on male faces is actually very high, it serves as an anchor upon which to build (1) the good accuracy of faces detected as male, by construction, and thus (2) the good accuracy of the correction on what is not detected as male. In this sense, 
the good accuracy on faces detection as male  
ensures that a correction based on manual validation on faces detected as female would accurately redress estimations for both genders. 
}

\begin{table*}[!th]
\caption{\label{tab:eval_models} {\bf Evaluation of the detection models.}}
\centering{
	\subfloat[Face detection][\label{tab:eval_face_detect}Face detection]{
	\begin{tabular}{cccc}
	\toprule
	&&\multicolumn{2}{c}{\bf Humans}\\
	&&\em Positive&\em Negative\\\midrule
	\multirow{2}{*}{\bf Model}&\em Positive&\cellcolor{MyGray}{977}&23\\
	&\em Negative&137&\cellcolor{MyGray}{{863}}\\
	\bottomrule
	\end{tabular}}\hspace{4em}
	\subfloat[Gender inference][\label{tab:eval_gender_detect}Gender inference]{
	\begin{tabular}{cccccc}
	\toprule
	&&\multicolumn{4}{c}{\bf Humans}\\
	&&\em Female&\em Male&\em Doubt&\em No face\\\midrule
	\multirow{2}{*}{\bf Model}&\em Female&\cellcolor{MyGray}{304}&162&18&16\\
	&\em Male&75&\cellcolor{MyGray}{410}&8&7\\
	\bottomrule
	\end{tabular}}
	}
\end{table*}

Thanks to this contextual validation step, we can now correct inference results appropriately. 
Knowing the shape and magnitude of model error makes it indeed easy to adjust face counts: for instance, if the algorithm detects a female face, we count .65 female faces and .35 male faces, using the confusion matrix of Table~\ref{tab:eval_models}. The same applies for male faces. In a nutshell, we adjust the raw \ffr{} using the following formula: $$\ffr{}_{\textrm{\footnotesize corrected}}=(1-\lambda)+(\lambda+\lambda'-1)\ffr{}$$ where $\lambda$ and $\lambda'$ are the proportions of true positives for male and female faces, respectively.
Furthermore,
we observe that algorithm error is not constant across time: female faces are over-estimated significantly more for the earlier than for the later years. In practice, we thus use time-dependent correction factors $\lambda$ and $\lambda'$ (based on time periods defined below for the {temporal} analysis).

\section*{Women's presence and its evolution}

The content analysis literature has relied on {diverse} features to assess gender representation in media{. It variously} mixed field expertise, subjective perceptions {and quantifiable variables.} These endeavors often led to semantic {characterizations} such as women appearing ``as dependent on men'', ``unintelligent'', ``less competitive'', ``more sexualized''~\citep{busby1975sex}, which are identified, annotated and counted throughout the media for further comment. The more formal the feature, the easier it is to scale the analysis to more observations, either by increasing the number of observers or automating the process.

\begin{figure}
	\centering
	\includegraphics[width=\linewidth]{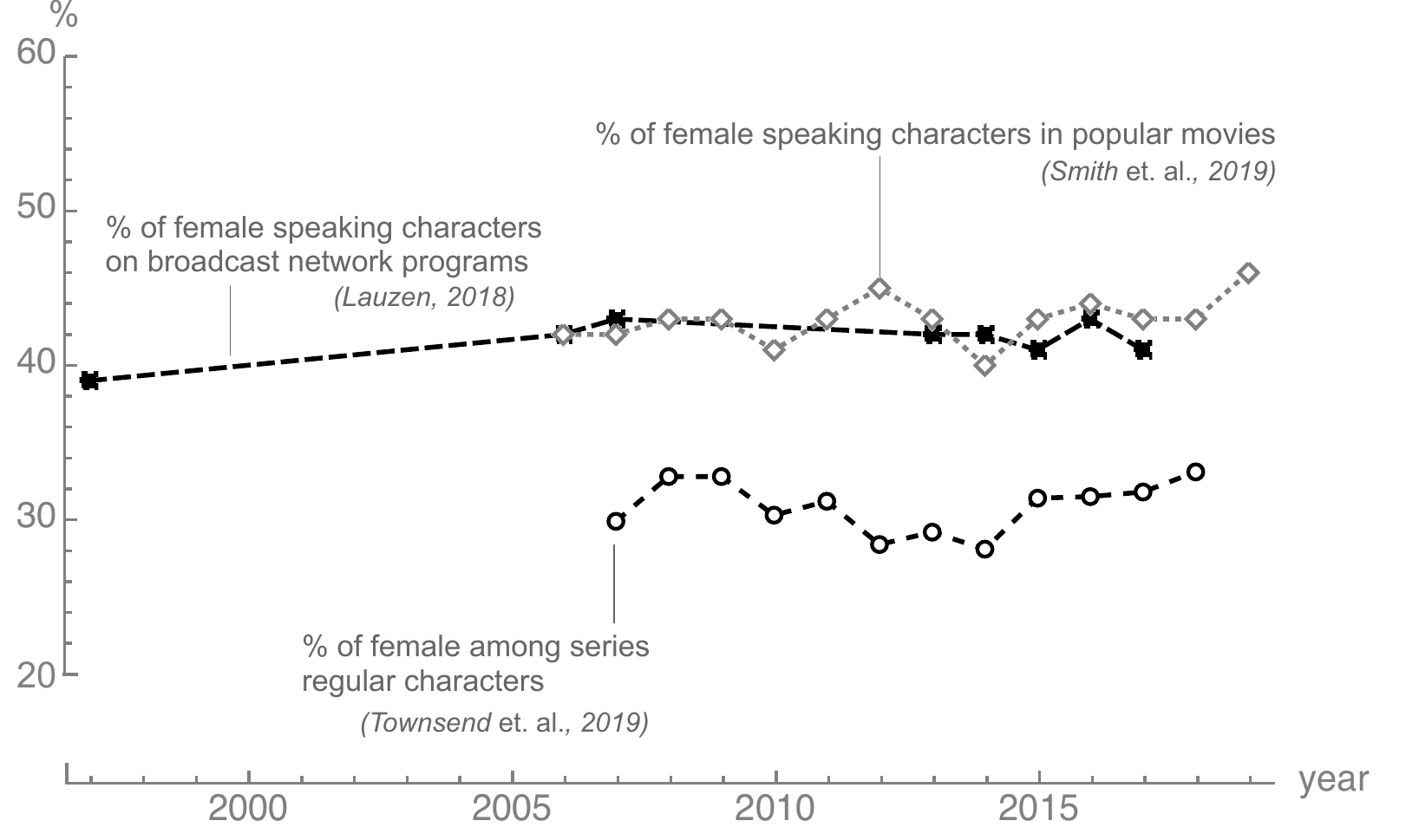}
	\caption{{\bf Several metrics used in the literature, based on \cite{smith2019inequality,lauzen2018boxed,glaad2019tv}}}
	\label{fig:biblio_ratios}
\end{figure}

More recently, various academic and activist projects have undertaken large scale analysis of visual entertainment media.  They often {lessened} the semantic complexity of the variables they rely on {and mainly focused on presence ratios}, while {being able to increase} sample sizes to a point that made {temporal} analysis possible. Figure~\ref{fig:biblio_ratios} gathers some results from three of these projects~\citep{glaad2019tv,smith2019inequality,lauzen2018boxed}. {They} not only confirm the under-representation of women {already} widely observed across the literature~\citep{busby1975sex,collins2011content}, but {they} also {invite} the conclusion that this situation has not evolved markedly in any direction during the considered periods.

\subsection*{Female face ratio (\ffr)}

\begin{figure*}[!h]
	\centering
	\includegraphics[width=.7\linewidth]{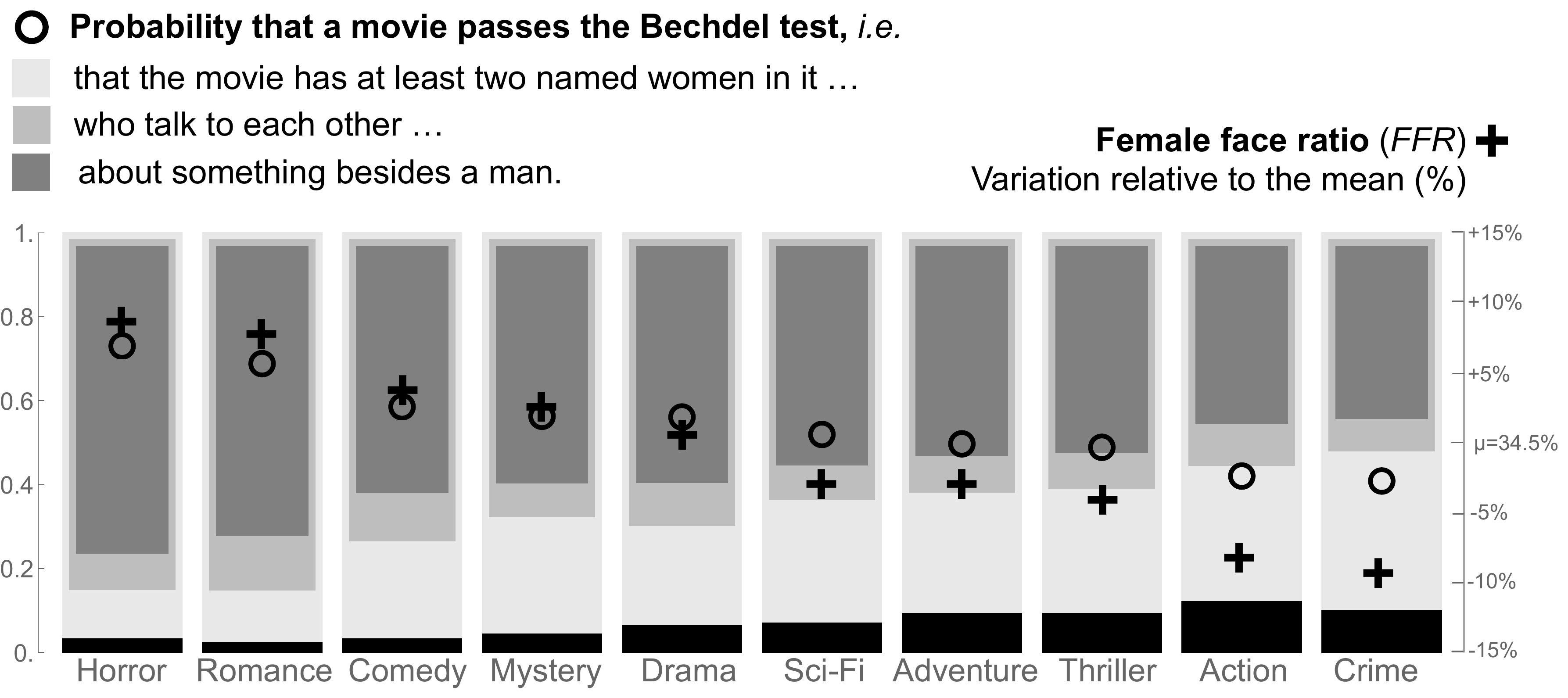}
	\caption{{\bf {Bechdel test and female face ratio (\textsc{ffr}) across a selection of popular movie genres}}.}
	\label{fig:bechdel_and_face_ratio}
\end{figure*}

{The face and gender detection algorithms we use provide us, for each movie frame, with three types of information of increasing complexity: number, gender and position of faces. In turn, we derive three types of variables. The first one is the most} minimalist: the percentage of faces classified as female among all the detected faces {on all frames of a given movie}, or \emph{female face ratio} (\ffr). 

{The average \ffr{} over all movies} is {$34.52\%$ ($\sigma=9.19$)}. This ratio {is comparable to what is} found in the literature, such as the ratio of female among characters in primetime television programming ($39.6\%$)~\citep{sink2017depictions} or among speaking characters in broadcast network programs and popular movies (see  Fig.~\ref{fig:biblio_ratios})~\citep{smith2019inequality,lauzen2018boxed}. However, {the \ffr} markedly differs from one genre to another: {we find} for example an average \textsc{ffr} of {$31.3\%$} for {\emph{Crime}} movies {while it reaches} {$37.1\%$ for \emph{Romance} movies.}

{To illustrate informally what the \textsc{ffr} {means in practice}, we {provide a few examples} of top grossing movies for some {domains} of this metric. First, among movies with a high percentage of male faces (\emph{i.e.} \textsc{ffr} $< 25\%$) we find {movies} such as \emph{Pirates of the Caribbean} (2007), \emph{Star Wars} (2005), \emph{Matrix} (2003), \emph{Independence Day} (1996) or \emph{Forest Gump} (1994), all with a \textsc{ffr} of around $23\%$. Movies such as \emph{The Hunger Games} (2014) and \emph{Jurassic World} (2015), \emph{Rogue One} (2016) and \emph{Gravity} (2013) lie around a female-male parity, with a \textsc{ffr} of between 45\% and 55\%. Lastly, the movie with the highest \ffr{} ($68\%$) is \emph{Bad Moms} (2016), closely followed by movies such as \emph{Sisters} (2015), \emph{Life of the Party} (2018) and \emph{Cake} (2014).}

Beyond these few examples, we further check how the \textsc{ffr} is correlated with narrative features {by comparing} it with the \cite{bechdel1983} test. This test is referenced and used in numerous studies~\citep{selisker2015bechdel,yang2020measuring,lindner2015million} and renowned for discarding around half of all reviewed movies with the simple criteria that two named women be present, speak to each other, about something besides a man. {We rely on} data {produced by} volunteers who manually evaluate if a movie passes or not the above cited conditions. This data is available at \href{https://bechdeltest.com/}{bechdeltest.com} and {only} covers a subset of our dataset ({n=$2,454$}). {As the \ffr{} varies along movie genres, so does the test}: we compared both metrics across the 10 most frequent movie genres, {as shown on} Figure~\ref{fig:bechdel_and_face_ratio}. {We find that} they are ordered in almost the same manner (Spearman score {$> 0.93$}) {even though the \ffr{} varies somewhat less across genres in absolute values}.

\paragraph{{{Temporal} analysis.}}

\begin{figure*}[!th]
	\begin{center}
		\subfloat[\label{fig:dist_evol_main}{Distributions of \ffr{} for each period.}]{\includegraphics[width=\linewidth]{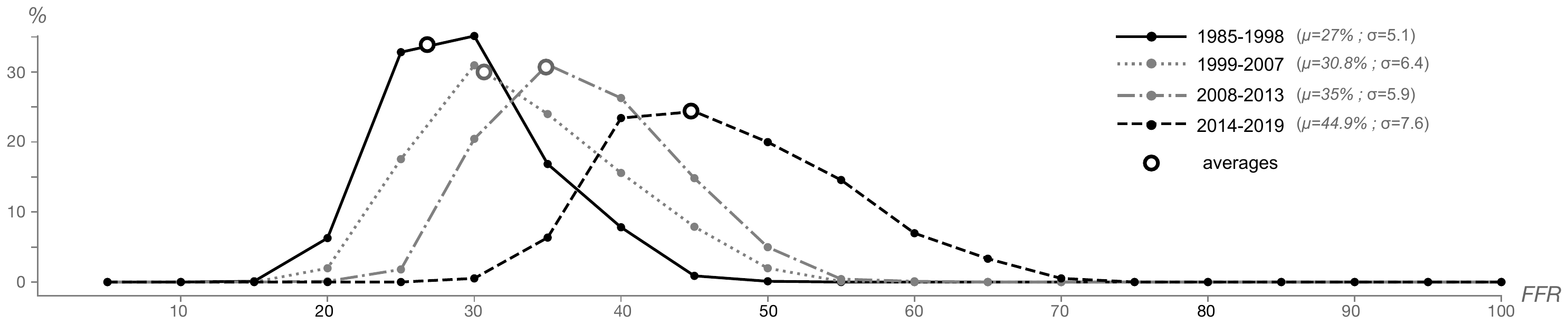}}
		\quad\quad
		\subfloat[\label{fig:dist_features}{Distributions of several features over the distribution of \ffr{}.}]{\includegraphics[width=\linewidth]{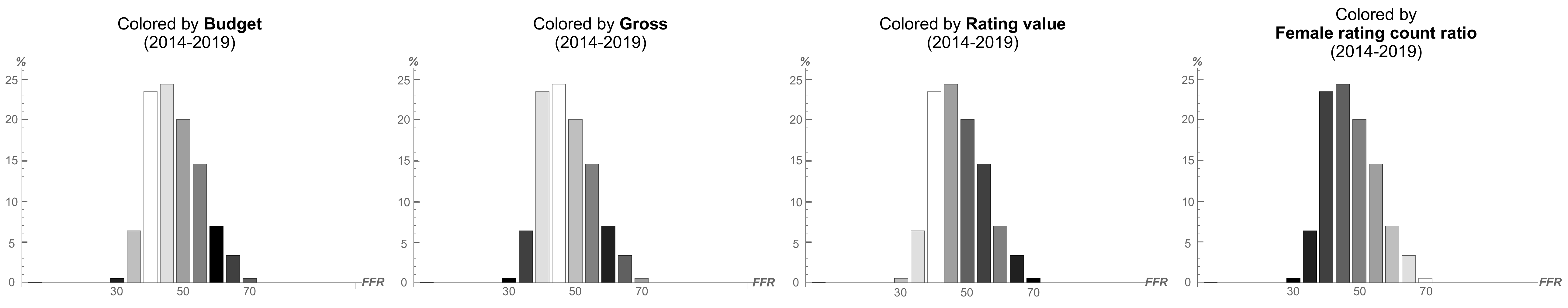}}
	\end{center}
	\caption{\textbf{Distributions of female face ratio} (\textsc{ffr}): {\bf (a)} \emph{Percentage of movies with a given \textsc{ffr}, one data point every $5$\%;}
	{\bf (b)} \emph{Percentage of movies with a given \textsc{ffr}, colored by the given variable mean within the bin, the lighter the higher.}}
\end{figure*}

Our aggregate findings on the \ffr{} since 1985 confirm women under-representation {in terms of on-screen presence. Yet, they also show} a significant trend toward {less inequality}. Our computational approach enables us to go into more detail by providing a relatively high resolution on the \ffr{} distribution across the observation period which, in turns, reveals several features.

We temporally divided our dataset into {quartiles, \hbox{i.e.}} four consecutive periods featuring the same number of films. 
As shown in Figure~\ref{fig:dist_evol_main}, the \ffr{} markedly increase{s across time} from an average $27\%$ 
between 1985 and 1998 to a mean \ffr{} of $44.9\%$ for the last period (2014-2019), close to a female-male balance. {The evolution of \ffr{} ranges is equally significant: most movies shot over 1985-1998 exhibit an \ffr{} of 20-45\%, while  movies of the most recent period 2014-19 generally cover the 35-65\% range. Besides}, the standard deviations of the underlying distributions increase overall (from $5.1$  to $7.6$). {This} probably {denotes} a higher {diversity} of {situations with regard to on-screen} gender {presence}. {On the whole, it} 
seems to be slowly evolving in favor of female representation as distributions appear to be increasingly right-skewed, \hbox{i.e.} towards a higher \ffr{}.
{Furthermore, considering data from \url{bechdeltest.com} restrained to the films of our datasets, over the same periods, we also observe an increase in the percentage of movies passing the test: 51\% between 1985 and 1998 up to 60\% for the last period (2014-2019). This evolution is comparable to the increase of the \ffr{}, albeit of a somewhat smaller magnitude -- +9\% \hbox{vs.} +18\%.}

As previously mentioned, while the literature widely acknowledges that women are under-represented in movies and, more broadly, in visual entertainment media, it usually states that this situation does not exhibit any significant evolution (see Fig~\ref{fig:biblio_ratios}).
{As it stands, we observe on our dataset a positive evolution over time of two distinct features, the \ffr{} and the Bechdel test success probability, in apparent contradiction with the hitherto observed stable representation of women.} {Note however that we exhibit a correlation between the \ffr{} and the Bechdel test, indicating that the \ffr{} nonetheless captures at least in part some semantic features beyond the plain proportion of female faces.}

{We can think of two phenomena to explain the discrepancy between our study and the previous ones. The first one relates to the way we select content, whereby we focus on a selection of films that may be {distinct from} what is immediately available on prime-time TV and on-demand streaming platforms. {In other words, both {ours} and the Bechdel test data are based on information contributed by users (on such and such website, relating to the interest of users for such and such content), while the traditional data is based on top-grossing films and/or programs (indicating what is offered to, or most successful for a given audience)}. The second one may be linked to the potential difference between on-screen presence (that we measure here) and more sophisticated features, such as effective speaking time or regularity of appearance {(that is typically measured in the literature)}.}

In essence, {the discrepancy} may demonstrate that there has been a significant {evolution} towards more {on-screen female presence} close to reaching female-male parity, but that this trend is {only moderately related to} the actual importance or influence of women in popular movies and their scenarios. In other words, {put in perspective with the literature, the evolution that we uncover here may not be of sufficient influence} on gender representation in popular movies. Figuring {out} {to what degree} {the increase of female on-screen presence} is {potentially preludial to an upcoming fairer gender representation}, or a {subtle} expression of {``purplewashing''}, would require a deeper qualitative analysis which is beyond the scope of our study.  

\paragraph{{Relation between \ffr{} and audience.}}
{We could see that distinct genres correspond to differing \ffr{} values. Budget and audience-related metadata enable us to characterize more finely the type of films that correspond to certain areas of the \ffr{} distribution. In Figure~\ref{fig:dist_features} we focus on \ffr{} histogram for the most recent period (2014-19). On this histogram, we project the average rank of movie \ffr{} with respect to} budget, gross, rating given by users (rating value) or number of people having rated a movie (rating count). 
Note that we chose to color histograms from white to black using rankings rather than absolute values, for there are wide variations in the orders of magnitude of the underlying average values (for instance, budget spans several orders of magnitude -- if a certain range of absolute values corresponded to a certain tone, we would almost have had either only white or only black bars, losing a significant resolution and missing the actual ordering and hierarchy between high-budget and low-budget movies).
Lighter tones correspond to higher ranks: for instance, the white bar for the ``budget'' coloration (left-most histogram) denotes the highest movie budgets. It coincides with the main mode and specifically with the bar of the histogram featuring the highest proportion of movies, with an \ffr{} of 35\%. The darkest tones, on the other hand, are found for the most extreme values of \ffr{} (very small or very high). Some exceptions are notable: there is a slightly {less dark} tone for \ffr{} {values} around 70\%, indicating the existence of relatively {higher} budget movies on that side as well. On the whole, the same phenomena are visible for world wide gross and rating. This suggests that the audience and their opinion resonate best with movies close to the main \ffr{} mode, which corresponds to the average \ffr{}  under-representation of women. Interestingly, the {higher \ffr{} values that emerged over the recent years (around 60\%) also correspond to relatively well-funded and successful movies. The last (right-most) histogram focuses on one of the best ordered tone scales (i.e., gray levels and \ffr{} values are ordered similarly), with respect to the proportion of user ratings given on IMDb by females. In other words, it reveals a virtually perfect agreement between movies featuring a high \ffr{} and the engagement of women in rating these movies (regardless of the polarity of these ratings, positive or negative).}

\section*{The framing of gender}

\subsection*{Face-ism}  
{From} an experimental psychology perspective, little is known about the effect upon observers of visual composition and element framing in a picture~\citep{sammartino2012aesthetic}. {A} movie shot composition allegedly helps convey emotional attachment of viewers to characters and narrative elements, driving them through the plot. These elements {have been} widely discussed and commented since the early research on modern aesthetics
{, including film theories~\citep{eisenstein1949film}}, and taken as basis for a more socio-political critique of public \emph{displays} of information such as gender~\citep{goffman1979gender}.
While the features extracted in the present study are insufficient for recovering the highly qualitative nature of such editorial choices, they still enable us to discuss character framing, of interest in film theory and its history~\citep{cutting2015framing}. In particular, by focusing on simple elements such as face position and surface, we first explore the hypothesis of \emph{face-ism} made by several gender studies. We further propose a more general appraisal of on-screen gender presence. 
This analysis is more sophisticated than the computation of \ffr{}: in particular, propagating the above-mentioned inference correction of the algorithm to complex on-screen face positions (bounding box areas) and compositions (one or several faces)  would prove to be quite arduous. For this reason and the sake of simplicity, we now restrict our analysis to the latest period of our dataset (2014-19), since model error was lowest and least serious. First, the accuracy of gender detection lies around 78\% and, more importantly, it is \emph{symmetric} across genders: male faces detected as female are in the same proportion as female faces detected as male.

\emph{Face-ism} is the tendency of an image to reveal more of the subject's face or head than body. It has been commonly associated with dominance and positive affect in audience perceptions~\citep{archer1983face}. Both in mass media and social networks~\citep{smith2012international}, research tends to observe that higher face-ism is granted to males over females.

In our dataset, the area of the face occupied on-screen can be assessed by the area of the face's bounding box. Compared with the size of the frame for a given movie, it yields the percentage of the frame occupied by a detected face, which can then be compared between movies with different aspect ratio or resolution. The values of face areas across all our dataset follows a heterogeneous distribution (technically a power law: many are small, few are large) with $80$\% faces occupying more that $1.36$\% of the frame. The median face area is $3.8$\% of the frame and is almost identical for male and female faces. {More precisely, the differences are statistically significant according to the non-parametric Mann-Whitney U test, yet extremely small: male general face area median is $0.03$\% above female. Furthermore, by genre these small differences appear sometimes in one direction, sometimes in another -- on the whole a typical signature of an effect that rather fluctuates around zero with some certainty.} 
This tends to not confirm the presence of gender biases in the way face-ism is granted to a character. Note however that our metric does not perfectly reflect potential face-ism, for it lacks the ability to compare the area of the face with that of the body -- caution must hence be applied before drawing from this result a refutation of the hypothesis of gender bias in face-ism. {Further automated inquiry on the matter should therefore make use of an additional algorithm able to detect and measure the presence of bodies in the picture.}

\subsection*{Gender's \emph{mise-en-scene} and \emph{mise-en-cadre}}

\begin{figure}[!hb]
	\centering
	\includegraphics[width=.8\linewidth]{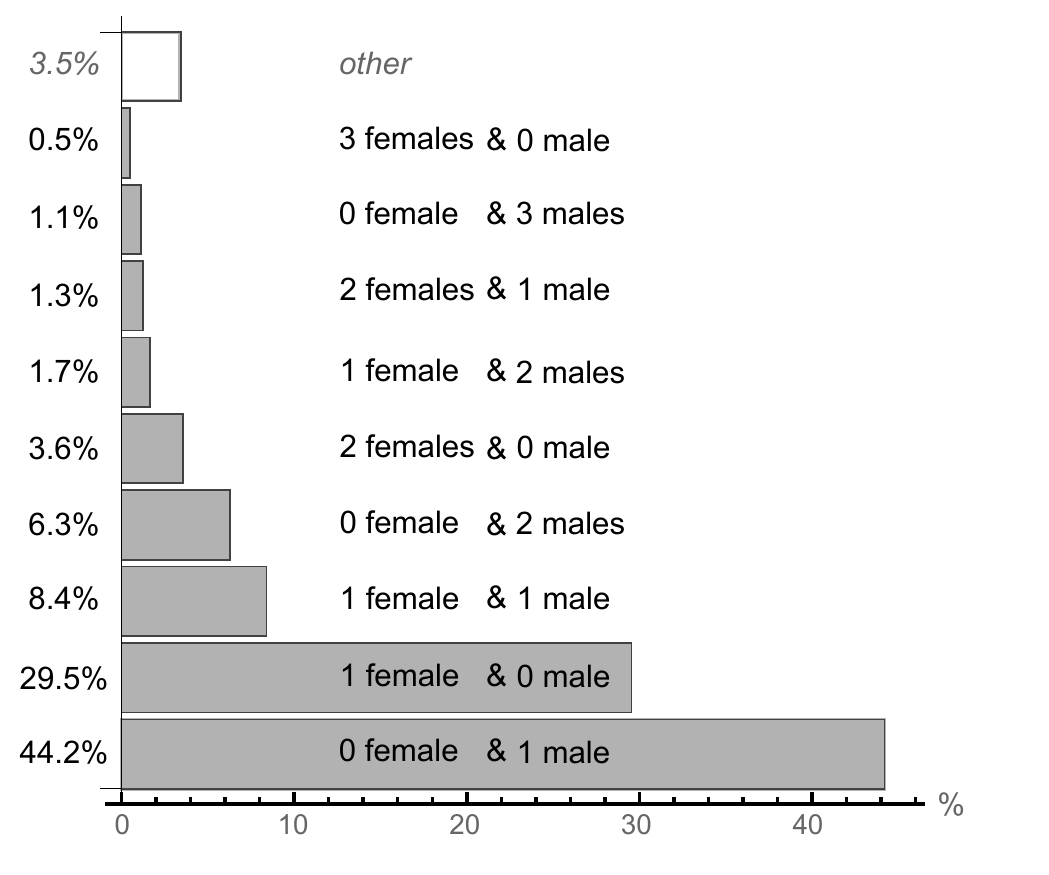}
	\caption{{{\bf Combinations of character gender} (2014-2019).}}
	\label{fig:gender_combi}
\end{figure}

Choosing how many characters appear in a given frame is an influential element of the craft of staging, or \emph{mise-en-scene}. It may direct the viewer's attention to one face or divide it among several, significantly modifying the {perception of actors'} performance and {their surroundings}. Thus, we analyzed the combinations of character genders appearing in a same frame. As shown in figure~\ref{fig:gender_combi}, the distribution of the most observed combinations reveals that {9} cases account for more that 95\% of all frames with faces and that the one-male-only configuration represents almost half of them. 

\begin{figure*}[!h]
	\centering
	\includegraphics[width=\linewidth]{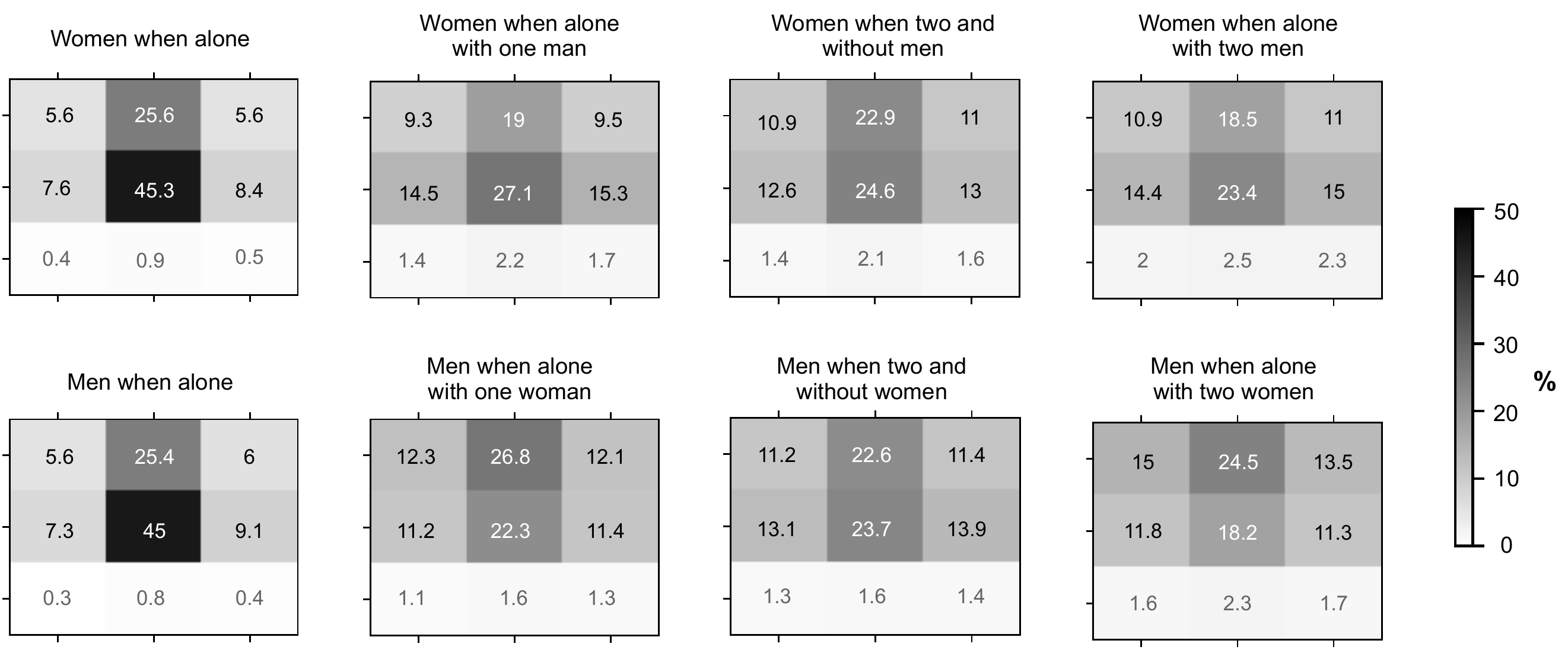}
	\caption{{{\bf Distribution of faces position on-screen} (2014-2019).}}
	\label{fig:mignonnes_matrices}
\end{figure*}

{Let us first focus on frames with only one face, which is} the most common case. {The distribution of the gender of that face exhibits a more marked bias in favor of male faces than the \ffr{}: 40\% of one-face frames feature a female, \hbox{vs.} 60\% for males (44.2\% out of 29.5+44.2\%), while the average global \ffr{} for the last period is 44.9\%. In other words, there seems to be a stronger bias favoring male presence in situations featuring a single face.}

{Furthermore}, following the ranking of figure~\ref{fig:gender_combi} in decreasing order exhibits a perfect symmetry of gender combinations (0 female/1 male, 1/0, 0/2, 2/0, etc), with equivalent configurations appearing first (\hbox{i.e.} 0 female/1 male before 1 female/0 male), in line with the underlying general bias in favor of male face presence. This hint at the idea that there is no significant additional gender bias in the character composition of a frame beyond the general previously observed 45-55 woman-man representation unbalance for that period.

We used these combinations to see if gender has an impact on the screen location of faces or, in other words, to observe if there is a gender-specific \emph{mise-en-cadre} depending on these configurations. Figure~\ref{fig:mignonnes_matrices} displays small matrices representing the screen on which a movie would be displayed, split according to the common rule-of-thirds. Each zone is annotated with the percentage of women or men appearing in it, in the context of the character gender combination mentioned above it.

We used chi-square to test the hypothesis of independence of the frequency distributions found in the various matrices. We considered the categorical variable \emph{mise-en-cadre}, with 9 possible values (one for each position in the 3x3 grid). We generated a contingency table for each pair of face configurations. We also checked for aggregated horizontal and vertical positions, in such cases the \emph{mise-en-cadre} only having 3 possible categories (in the horizontal case: left, center, right, in the vertical case: top, middle, bottom). For all these cases and all pairwise combinations we found strong support for \emph{dependence}, with all p-values $< 0.005$. This leads us to conclude that even differences of small magnitude are statistically significant.

When in a gender-mixed configuration, women are more present in the middle third of the screen while men seem to appear more frequently in the upper third of the screen. A similar phenomenon can be observed when women and men are alone or in a non-mixed character gender combination, but in these cases, while the observation is still statistically significant, the magnitude of the effect is very small. 

\begin{figure}[!h]
	\centering
	\includegraphics[width=\linewidth]{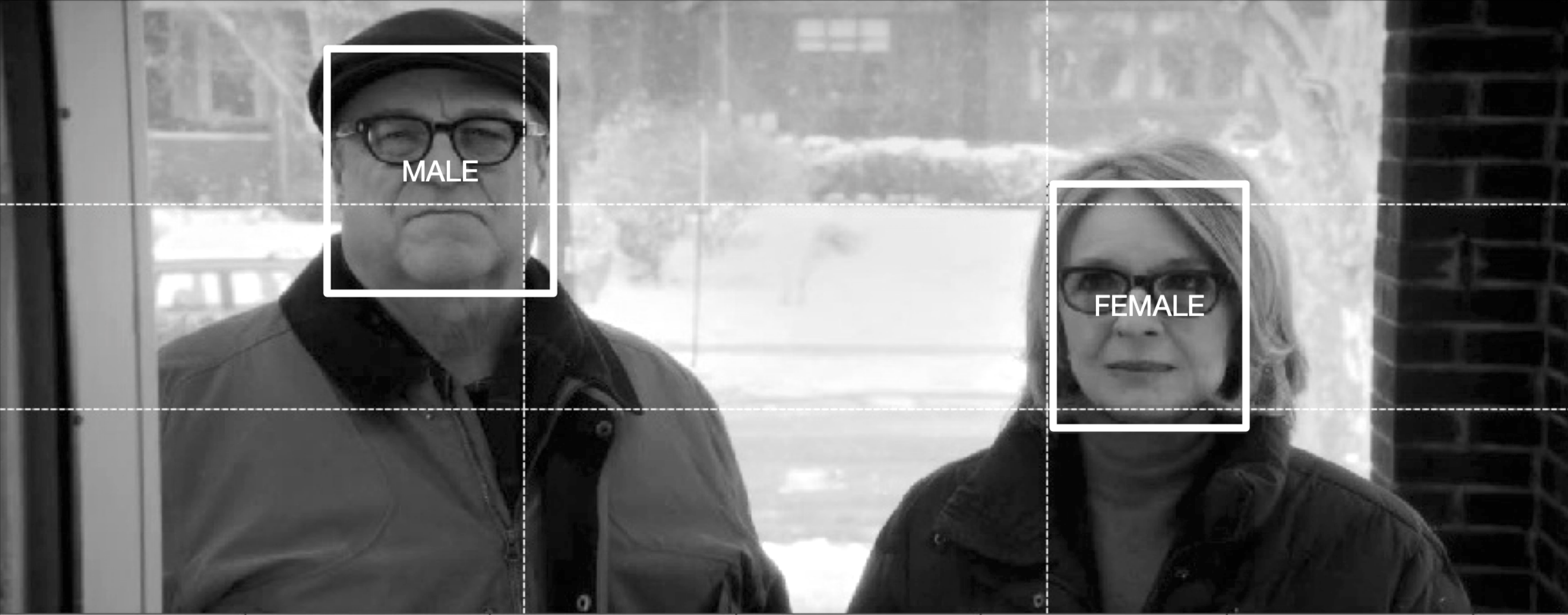}
	\caption{{\bf {Example of gender placement on-screen.}}
	}
	\label{fig:example_rule_third}
\end{figure}

{We randomly selected hundreds of pictures exhibiting this significant pattern: the woman's face present in the middle third of the screen while the man's is located in the upper third. A {manual evaluation} of this selection revealed that this bias is partly due the height gap between actors and actresses, as illustrated by Figure~\ref{fig:example_rule_third}. As stated earlier, the \emph{mise-en-cadre} of characters goes beyond face size and position. A more fine-grained analysis would require the ability to assess subtle biases in depth and perspective of characters placement together explanatory and evaluation protocols with movies experts. {We leave this to further research.}}

\section*{Concluding remarks}

In practice, our contribution principally exhibits several gender representation discrepancies in on-screen presence in a large set of movies spanning a wide period of time. More broadly, this article also aims at demonstrating the usefulness and feasibility of automated computational methods for the study of gender representativeness in mass media. We successfully uncovered clear historical trends thanks to the possibility of handily producing empirical observations at a scale that would have been both expensive and difficult for a qualitative endeavor.
Nonetheless, our essentially quantitative approach did not prevent us to appraise more sophisticated features and to correlate our findings with a variety of meta-data. As such, our approach could be easily replicated on other corpuses within the visual entertainment industry, such as advertisement and TV shows. 

Meanwhile, our study also outlined several challenges for computational methods to efficiently tackle issues related to gender representation in media. Firstly, even though we used face and gender detection algorithms with solid track records from an engineering perspective, we had to realize and acknowledge that the underlying machine learning models still suffer from important and significant biases, especially with respect to the empirical context of movie content over several decades. Trusting the output of these algorithms at face value would have led to significant errors. 
The development of a protocol to assess their bias on a case-by-case basis proved to be key: further studies should imperatively estimate the performance of such tools, be it in the framework of gender studies or more broadly in the prospect of carrying out the ``distant viewing'' of media material.
Secondly, our results have shown clear trends towards more representativeness of on-screen woman presence in popular movies, whereas parts of the state of the art rather tend to report a rather stable (under-)representation. This opens up interesting venues for further qual-quant analyses: for instance, by focusing on movies quantitatively featuring a gender ratio close to parity and describing qualitatively how women are actually represented with respect to men. On the whole, we hope to have shown that there is a promising potential in the fine qualitative analysis of media material selected on the basis of a large-scale scanning of sizable media datasets.


\small

\subsection*{Data availability}

The datasets generated during and/or analysed during the current study are available in the Nakala repository, \url{https://doi.org/10.34847/nkl.543czc59}.

\subsection*{Acknowledgments}

The authors are grateful to Élise Marsicano, Lilas Duvernois, Cécile Dumas, Jean-Christophe Ribot and Angela Crone for their help and advices in conducting this research.

\subsection*{Competing interests}

The authors have declared that no competing interests exist.

\bibliographystyle{apalike}

\begin{thebibliography}{}

\bibitem[Archer et~al., 1983]{archer1983face}
Archer, D., Iritani, B., Kimes, D.~D., and Barrios, M. (1983).
\newblock Face-ism: Five studies of sex differences in facial prominence.
\newblock {\em Journal of Personality and social Psychology}, 45(4):725--735.

\bibitem[Arnold and Tilton, 2019]{arnold2019distant}
Arnold, T. and Tilton, L. (2019).
\newblock {Distant viewing: analyzing large visual corpora}.
\newblock {\em Digital Scholarship in the Humanities}, 34(S1):i3--i16.

\bibitem[Bechdel, 1983]{bechdel1983}
Bechdel, A. (1983).
\newblock {\em Dykes to Watch Out For}.
\newblock \url{https://dykestowatchoutfor.com/}.

\bibitem[Bost et~al., 2016]{bost2016narrative}
Bost, X., Labatut, V., Gueye, S., and Linar{\`e}s, G. (2016).
\newblock Narrative smoothing: dynamic conversational network for the analysis
  of tv series plots.
\newblock In {\em 2016 IEEE/ACM International Conference on Advances in Social
  Networks Analysis and Mining (ASONAM)}, pages 1111--1118. IEEE.

\bibitem[Buolamwini and Gebru, 2018]{buolamwini2018gender}
Buolamwini, J. and Gebru, T. (2018).
\newblock Gender shades: Intersectional accuracy disparities in commercial
  gender classification.
\newblock In {\em Conference on fairness, accountability and transparency},
  pages 77--91.

\bibitem[Busby, 1975]{busby1975sex}
Busby, L.~J. (1975).
\newblock Sex-role research on the mass media.
\newblock {\em Journal of Communication}, 25(4):107--131.

\bibitem[Chaney and Blei, 2012]{chaney2012visualizing}
Chaney, A. J.-B. and Blei, D.~M. (2012).
\newblock Visualizing topic models.
\newblock In {\em Proc. 6th ICWSM AAAI Conf on weblogs and social media}, pages
  419--422.

\bibitem[Cillessen and Marks, 2011]{cillessen2011conceptualizing}
Cillessen, A.~H. and Marks, P.~E. (2011).
\newblock Conceptualizing and measuring popularity.
\newblock {\em Popularity in the peer system}, pages 25--56.

\bibitem[Cohen, 2003]{cohen2003incentives}
Cohen, B. (2003).
\newblock Incentives build robustness in bittorrent.
\newblock In {\em Workshop on Economics of Peer-to-Peer systems}, volume~6,
  pages 68--72.

\bibitem[Collins, 2011]{collins2011content}
Collins, R.~L. (2011).
\newblock Content analysis of gender roles in media.
\newblock {\em Sex roles}, 64:290--298.

\bibitem[Crawford and Paglen, 2019]{crawford2019excavating}
Crawford, K. and Paglen, T. (2019).
\newblock Excavating ai: The politics of images in machine learning training
  sets.
\newblock \url{https://www.excavating.ai/}.

\bibitem[Cutting, 2015]{cutting2015framing}
Cutting, J.~E. (2015).
\newblock The framing of characters in popular movies.
\newblock {\em Art \& Perception}, 3(2):191--212.

\bibitem[Cutting and Candan, 2015]{ShotDurations}
Cutting, J.~E. and Candan, A. (2015).
\newblock Shot durations, shot classes, and the increased pace of popular
  movies.
\newblock {\em Projections}, 9(2):40--62.

\bibitem[Deng et~al., 2009]{deng2009imagenet}
Deng, J., Dong, W., Socher, R., Li, L.-J., Li, K., and Fei-Fei, L. (2009).
\newblock Imagenet: A large-scale hierarchical image database.
\newblock In {\em Proc. 2009 IEEE conference on computer vision and pattern
  recognition}, pages 248--255.

\bibitem[Dhomne et~al., 2018]{dhomne2018gender}
Dhomne, A., Kumar, R., and Bhan, V. (2018).
\newblock Gender recognition through face using deep learning.
\newblock {\em Procedia Computer Science}, 132:2--10.

\bibitem[Eisenstein, 1949]{eisenstein1949film}
Eisenstein, S. (1949).
\newblock {\em Film form: Essays in film theory}.
\newblock Harcourt, Inc.

\bibitem[Follows, 2014]{follows2014gender}
Follows, S. (2014).
\newblock Gender within film crews.
\newblock {\em Stephen Follows Film Data and Education}, 22.

\bibitem[Goffman, 1979]{goffman1979gender}
Goffman, E. (1979).
\newblock {\em Gender advertisements}.
\newblock Macmillan International Higher Education.

\bibitem[Guha et~al., 2015a]{guha2015gender}
Guha, T., Huang, C.-W., Kumar, N., Zhu, Y., and Narayanan, S.~S. (2015a).
\newblock Gender representation in cinematic content: A multimodal approach.
\newblock In {\em Proceedings of the 2015 ACM on International Conference on
  Multimodal Interaction}, pages 31--34.

\bibitem[Guha et~al., 2015b]{guha2015computationally}
Guha, T., Kumar, N., Narayanan, S.~S., and Smith, S.~L. (2015b).
\newblock Computationally deconstructing movie narratives: an informatics
  approach.
\newblock In {\em 2015 IEEE international conference on acoustics, speech and
  signal processing (ICASSP)}, pages 2264--2268. IEEE.

\bibitem[Guo and Zhang, 2019]{guo2019survey}
Guo, G. and Zhang, N. (2019).
\newblock A survey on deep learning based face recognition.
\newblock {\em Computer Vision and Image Understanding}, 189:102805.

\bibitem[Hassan et~al., 2018]{hassan2018achieving}
Hassan, H., Aue, A., Chen, C., Chowdhary, V., Clark, J., Federmann, C., Huang,
  X., Junczys-Dowmunt, M., Lewis, W., Li, M., Liu, S., Liu, T.-Y., Luo, R.,
  Menezes, A., Qin, T., Seide, F., Tan, X., Tian, F., Wu, L., Wu, S., Xia, Y.,
  Zhang, D., Zhang, Z., and Zhou, M. (2018).
\newblock Achieving human parity on automatic {C}hinese to {E}nglish news
  translation.
\newblock {\em arXiv}, 1803.05567.

\bibitem[Jang et~al., 2019]{jang2019quantification}
Jang, J.~Y., Lee, S., and Lee, B. (2019).
\newblock Quantification of gender representation bias in commercial films
  based on image analysis.
\newblock {\em Proceedings of the ACM on Human-Computer Interaction},
  3(CSCW):1--29.

\bibitem[Kataria and Kumar, 2016]{Kataria2016SceneIE}
Kataria, S. and Kumar, A. (2016).
\newblock Scene intensity estimation and ranking for movie scenes through
  direct content analysis.
\newblock Project report, IIT Kanpur.

\bibitem[Kian et~al., 2009]{kian-espn-2009}
Kian, E. T.~M., Mondello, M., and Vincent, J. (2009).
\newblock Espn---the women's sports network? a content analysis of internet
  coverage of march madness.
\newblock {\em Journal of Broadcasting \& Electronic Media}, 53(3):477--495.

\bibitem[Ko et~al., 2019]{ko2019learning}
Ko, M.-Y., Li, J.-L., and Lee, C.-C. (2019).
\newblock Learning minimal intra-genre multimodal embedding from trailer
  content and reactor expressions for box office prediction.
\newblock In {\em 2019 IEEE International Conference on Multimedia and Expo
  (ICME)}, pages 1804--1809. IEEE.

\bibitem[Lauzen, 2018]{lauzen2018boxed}
Lauzen, M.~M. (2018).
\newblock Boxed in 2017-18: Women on screen and behind the scenes in
  television.
\newblock {\em Center for the Study of Women in Television and Film, San Diego
  State University.}

\bibitem[Lauzen, 2019]{lauzen2019sa}
Lauzen, M.~M. (2019).
\newblock It's a man's (celluloid) world: Portrayals of female characters in
  the top grossing films of 2018.
\newblock {\em Center for the Study of Women in Television and Film, San Diego
  State University}.

\bibitem[Lindner et~al., 2015]{lindner2015million}
Lindner, A.~M., Lindquist, M., and Arnold, J. (2015).
\newblock Million dollar maybe? the effect of female presence in movies on box
  office returns.
\newblock {\em Sociological Inquiry}, 85(3):407--428.

\bibitem[Mani, 2001]{mani-auto}
Mani, I. (2001).
\newblock {\em Automatic Summarization}.
\newblock John Benjamins Publishing Company.

\bibitem[McBee et~al., 2018]{mcbee2018deep}
McBee, M.~P., Awan, O.~A., Colucci, A.~T., Ghobadi, C.~W., Kadom, N., Kansagra,
  A.~P., Tridandapani, S., and Auffermann, W.~F. (2018).
\newblock Deep learning in radiology.
\newblock {\em Academic radiology}, 25(11):1472--1480.

\bibitem[Moretti, 2000]{moretti-2000-conjectures}
Moretti, F. (2000).
\newblock Conjectures on world literature.
\newblock {\em New Left Review}, 1:54--68.

\bibitem[Neuendorf, 2017]{neuendorf2016content}
Neuendorf, K.~A. (2017).
\newblock {\em The content analysis guidebook}.
\newblock Sage.

\bibitem[Rafaeli and Ariel, 2008]{rafaeli2008online}
Rafaeli, S. and Ariel, Y. (2008).
\newblock Online motivational factors: Incentives for participation and
  contribution in wikipedia.
\newblock {\em Psychological aspects of cyberspace: Theory, research,
  applications}, 2(08):243--267.

\bibitem[Rudy et~al., 2010]{rudy2010context}
Rudy, R.~M., Popova, L., and Linz, D.~G. (2010).
\newblock The context of current content analysis of gender roles.
\newblock {\em Sex roles}, 62:705--720.

\bibitem[Sammartino and Palmer, 2012]{sammartino2012aesthetic}
Sammartino, J. and Palmer, S.~E. (2012).
\newblock Aesthetic issues in spatial composition: Effects of vertical position
  and perspective on framing single objects.
\newblock {\em Journal of Experimental Psychology: Human Perception and
  Performance}, 38(4):865.

\bibitem[Selisker, 2015]{selisker2015bechdel}
Selisker, S. (2015).
\newblock The bechdel test and the social form of character networks.
\newblock {\em New Literary History}, 46(3):505--523.

\bibitem[Sink and Mastro, 2017]{sink2017depictions}
Sink, A. and Mastro, D. (2017).
\newblock Depictions of gender on primetime television: A quantitative content
  analysis.
\newblock {\em Mass Communication and Society}, 20(1):3--22.

\bibitem[Smith and Cooley, 2012]{smith2012international}
Smith, L.~R. and Cooley, S.~C. (2012).
\newblock International faces: An analysis of self-inflicted face-ism in online
  profile pictures.
\newblock {\em Journal of Intercultural Communication Research},
  41(3):279--296.

\bibitem[Smith et~al., 2019]{smith2019inequality}
Smith, S.~L., Choueiti, M., Pieper, K., Yao, K., Case, A., and Choi, A. (2019).
\newblock {\em Inequality in 1,200 Popular Films}.
\newblock
  \url{http://assets.uscannenberg.org/docs/aii-inequality-report-2019-09-03.pdf}.

\bibitem[Somandepalli et~al., 2021]{somandepalli2021computational}
Somandepalli, K., Guha, T., Martinez, V.~R., Kumar, N., Adam, H., and
  Narayanan, S. (2021).
\newblock Computational media intelligence: Human-centered machine analysis of
  media.
\newblock {\em Proceedings of the IEEE}.

\bibitem[Townsend et~al., 2019]{glaad2019tv}
Townsend, M., Deerwater, R., Adams, N., Trasandes, M., and Hood, D. (2019).
\newblock {\em Where we are on TV}.
\newblock GLAAD.

\bibitem[Vassileva, 2002]{vassileva2002motivating}
Vassileva, J. (2002).
\newblock Motivating participation in peer to peer communities.
\newblock In {\em International Workshop on Engineering Societies in the Agents
  World}, pages 141--155. Springer.

\bibitem[Wolfram, 2020]{referencewolfram2020facialfeatures}
Wolfram (2020).
\newblock {FacialFeatures}.
\newblock \url{https://reference.wolfram.com/language/ref/FacialFeatures.html}.
\newblock Accessed: March 5, 2021.

\bibitem[Xu et~al., 2015]{xu2015show}
Xu, K., Ba, J., Kiros, R., Cho, K., Courville, A., Salakhudinov, R., Zemel, R.,
  and Bengio, Y. (2015).
\newblock Show, attend and tell: Neural image caption generation with visual
  attention.
\newblock In {\em International conference on machine learning}, pages
  2048--2057.

\bibitem[Yang and Lai, 2010]{yang2010motivations}
Yang, H.-L. and Lai, C.-Y. (2010).
\newblock Motivations of wikipedia content contributors.
\newblock {\em Computers in human behavior}, 26(6):1377--1383.

\bibitem[Yang et~al., 2020]{yang2020measuring}
Yang, L., Xu, Z., and Luo, J. (2020).
\newblock Measuring women representation and impact in films over time.
\newblock {\em arXiv preprint arXiv:2001.03513}.

\bibitem[Zech et~al., 2018]{zech2018confounding}
Zech, J.~R., Badgeley, M.~A., Liu, M., Costa, A.~B., Titano, J.~J., and
  Oermann, E.~K. (2018).
\newblock Confounding variables can degrade generalization performance of
  radiological deep learning models.
\newblock {\em arXiv preprint arXiv:1807.00431}.

\end{thebibliography}

\end{document}